\documentclass[11pt,a4paper]{article}

\usepackage{latexsym}
\usepackage{amssymb}
\usepackage{amsmath}
\usepackage{amsfonts}
\usepackage{mathrsfs}
\usepackage{cite}
\usepackage{bm}

\newcommand{\be}{\begin{equation}}
\newcommand{\ee}{\end{equation}}

\newcommand{\lb}{\label}

\def\tr{{\rm tr}}
\def\tr{{\rm tr}\,}
\def\Tr{{\rm Tr}\,}
\def\cN{{\cal N}}

\def\bea{\begin{eqnarray}}
\def\eea{\end{eqnarray}}
\def\nn{\nonumber}

\def\cN{{\cal N}}

\def\f{\frac}

\def\tr{{\rm tr}\,}

\def\nn{\nonumber}

\def\d{\delta}

\def\g{\gamma}

\def\ve{\varepsilon}

\def\sB{\stackrel{\frown}{\bm{\square}}}

\def\eq{\eqref}
\def\pr{\partial}
\def\nb{\nabla}

\numberwithin{equation}{section}

\begin{document}
%%%%%%%%%%%%%%%%%%%%%%%%%%%%%%%%%%%%%%%%%%%%%%%%%%%%%%%%%%%%%%%%%%
%\begin{titlepage}
\begin{center}
\vspace{1cm} {\Large\bf On gauge dependence of the one-loop
divergences
\vspace{0.1cm}

in $6D$, $\cN= (1,0)$ and $\cN= (1,1)$ SYM theories}
\vspace{1.5cm}

 {\bf
 I.L. Buchbinder\footnote{joseph@tspu.edu.ru }$^{\,a,b,c}$,
 E.A. Ivanov\footnote{eivanov@theor.jinr.ru}$^{\,c}$,
 B.S. Merzlikin\footnote{merzlikin@tspu.edu.ru}$^{\,a,c}$,
 K.V. Stepanyantz\footnote{stepan@m9com.ru}$^{\,d,c}$
 }
\vspace{0.4cm}

 {\it $^a$ Department of Theoretical Physics, Tomsk State Pedagogical
 University,\\ 634061, Tomsk,  Russia \\ \vskip 0.15cm
 $^b$ National Research Tomsk State University, 634050, Tomsk, Russia \\ \vskip 0.1cm
 $^c$ Bogoliubov Laboratory of Theoretical Physics, JINR, 141980 Dubna, Moscow region,
 Russia \\ \vskip 0.1cm
 $^d$ Department of Theoretical Physics, Moscow State University,
119991, Moscow, Russia }
\end{center}
\vspace{0.4cm}

\begin{abstract}
We study the gauge dependence of one-loop divergences in a general
matter-coupled $6D$,  $\cN=(1,0)$ supersymmetric gauge theory in the
harmonic superspace formulation. Our analysis is based on the
effective action constructed by the background superfield  method,
with the gauge-fixing term involving one real parameter $\xi_0$. A
manifestly gauge invariant and $\cN=(1,0)$ supersymmetric procedure
for calculating the one-loop effective action is developed. It
yields the one-loop divergences in an explicit form and allows one
to investigate their gauge dependence. As compared to the minimal
gauge, $\xi_0=1$, the divergent part of the  general-gauge effective
action contains a new  term depending on $\xi_0\,$. This term
vanishes for the background superfields satisfying the classical
equations of motion, so that the $S$-matrix divergences are
gauge-independent. In the case of $6D$, $\cN= (1,1)$ SYM theory we
demonstrate that some divergent contributions in the non-minimal
gauges  do not vanish off shell, as opposed to the minimal gauge.
\end{abstract}
%\end{titlepage}
%%%%%%%%%%%%%%%%%%%%%%%%%%%%%%%%%%%%%%%%%%%%%%%%%%%%%%%%%%%%%%%%%%
\setcounter{footnote}{0} \setcounter{page}{1}
%%%%%%%%%%%%%%%%%%%%%%%%%%%%%%%%%%%%%%%%%%%%%%%%%%%%%%%%%%%%%%%%%%

%%%%%%%%%%%%%%%%%%%%%%%%%%%%%%%%%%%%%%%%%%%%%%%%%%%%%%%%%%%%%%%%%%
\section{Introduction}
%\hspace*{\parindent}
The study of quantum aspects of the higher-dimensional supersymmetric gauge
field theories attracts a wide interest for a long time, mainly because of their use
for the low-energy description of diverse sectors of superstring theory (see, e.g., \cite{Giveon:1999px,Bagger:2012jb}). From the
field-theoretical point of view, such theories possess a rather unusual UV
behavior. Although they are non-renormalizable by
power counting, the relevant amplitudes can be still finite for some low numbers of loops.
In particular, in the six-dimensional maximally extended ${\cal
N}=(1,1)$ super Yang-Mills theory the one- and two-loop amplitudes are finite \cite{Fradkin:1982kf,Marcus:1983bd,Marcus:1984ei,Howe:1983jm,Howe:2002ui,Bossard:2009sy,Bossard:2009mn,Kazakov:2002jd,Bork:2015zaa}.

Recently, using techniques of the six-dimensional ($6D$) ${\cal
N}=(1,0)$ harmonic superspace \cite{Howe:1985ar,Zupnik:1986da,Bossard:2015dva} (which is a direct generalization of $4D$, ${\cal N}=2$
harmonic superspace \cite{Galperin:1985ec,Galperin:1984av,Galperin:2001uw}), we studied the quantum structure of ${\cal N}=(1,0)$ and ${\cal N}=(1,1)$
supersymmetric $6D$ gauge theories \cite{Buchbinder:2016gmc,Buchbinder:2016url,Buchbinder:2017ozh,Buchbinder:2017gbs,Buchbinder:2017xjb} (see also the review
\cite{Buchbinder:2018bhs}), paying a special attention to off-shell
divergences. We considered $6D, \, {\cal N}=(1,0)$
non-abelian Yang-Mills theory coupled to the hypermultiplet in an arbitrary
representation of the gauge group and calculated the one-loop
divergences of the superfield effective action. It was shown that, in the particular case of
${\cal N}=(1,1)$ theory, {\it i.e.} with the hypermultiplet in the adjoint
representation, all one-loop divergences vanish off shell. The
calculations were performed in the minimal gauge, in which the
gauge superfield propagator has the simplest form. The natural question to be posed was as to whether the vanishing of the
off-shell one-loop divergences  depends on the choice of the gauge-fixing condition.

We started studying the gauge dependence of the one-loop divergences in  $6D$, $\cN=(1,0)$
supersymmetric gauge theories in our previous work \cite{Buchbinder:2018lbd}.
As a simplest example of a supersymmetric theory, where the problem of
the gauge dependence occurs, we considered the abelian $\cN=(1,0)$
gauge theory and investigated the structure of the
gauge-dependent divergent contributions to the one-loop effective
action. It was shown that the one-loop divergences are actually
gauge-dependent. The present paper generalizes this study to the generic non-abelian $\cN=(1,0)$ SYM theory interacting
with a set of hypermultiplets in an arbitrary representation of the gauge
group. We consider the gauge conditions
involving one arbitrary gauge parameter $\xi_0$ (analogs of usual $\xi$-gauges in the non-supersymmetric case)
and analyze the dependence of the one-loop divergences on this parameter.

The analysis of the effective action in refs. \cite{Buchbinder:2016gmc,Buchbinder:2016url,Buchbinder:2017ozh,Buchbinder:2017gbs,Buchbinder:2017xjb} was
carried out in the framework of the harmonic superfield background field
method, which ensures both the classical gauge
invariance and $\cN=(1,0)$ supersymmetry as manifest symmetries. This method was earlier developed  for the minimal
gauge only\footnote{$6D$, ${\cal N}=(1,0)$ background field method is
a generalization of $4D, {\cal N}=2$ background field method
worked out in \cite{Buchbinder:1997ya}. For a review of its various applications
see ref. \cite{Buchbinder:2016wng}.}. Now we formulate the
background superfield method for the one-parametric
family of the gauge conditions. The detailed
analysis of the structure of divergences in the non-minimal gauges
can hopefully be useful for a better understanding of the UV behavior of the theory
under consideration.

The paper is organized as follows. Sect. \ref{Section_Model}  provides some
basic knowledge about $6D$, ${\cal N}=(1,0)$ SYM theory interacting
with hypermultiplets in the harmonic superspace formulation. We briefly discuss
the superfield content and the action of the model. In Sect. \ref{Section_Effective_Action}
we develop the background superfield method with the gauge-fixing term containing an arbitrary real parameter and derive the
formal expression for the corresponding one-loop effective
action. In Sect. \ref{Section_Divergences} we calculate the one-loop divergences as functions of the gauge-fixing parameter. We
explicitly demonstrate that the divergences are gauge-dependent for both ${\cal N}=(1,0)$ and ${\cal N}=(1,1)$
theories\footnote{It is worth pointing out that $6D$, ${\cal
N}=(1,0)$ gauge theories are in general anomalous \cite{Townsend:1983ana,Smilga:2006ax,Kuzenko:2015xiz,Kuzenko:2017xgh}.
However, while investigating the one-loop divergences, the anomalies do not matter.}. We show that the gauge dependence of the divergences vanishes for
the background superfields satisfying the classical equations
of motion. This implies that the divergences of $S$-matrix are gauge-independent, in agreement with the general theorems
(see, e.g., \cite{DeWitt:1965jb})\footnote{The gauge dependence of
the one-loop divergences in $\cN=(1,0)$ SYM theory was also
discussed in ref. \cite{Kazakov:2002jd} in the framework of the component
formulation.}. We also discuss the one-loop divergences for the case of $\cN=(1,1)$ SYM theory, that is
$\cN=(1,0)$ SYM theory minimally coupled to a hypermultiplet in the adjoint representation of the gauge group. Sect. \ref{Summary} contains a summary of
the results obtained and a proposal for further work.

\section{Basic notions}
%\hspace*{\parindent}
\label{Section_Model}
The harmonic $6D$, $\cN=(1,0)$ superspace in the central basis is parametrized by the coordinates $(z, u) \equiv (x^M, \theta^a_i, u^{\pm
i})$. Here $x^M$, $M= 0,..,5$, are $6D$ Minkowski space-time
coordinates, $\theta^a_i$, $a=1,..,4\,$, $i=1,2\,$, are Grassmann
variables, and the additional harmonic variables $u^{\pm}_i$, $u^{+ i}u^-_i =1\,$, represent the coset $SU(2)/U(1)\,$,  with $SU(2)$ being $R$-symmetry group
of $6D$, $\cN=(1,0)$ Poincar\'e superalgebra \cite{Howe:1985ar,Zupnik:1986da}.

The harmonic superspace in the analytic basis is parametrized, along with the harmonic variables, by
the analytic coordinate $z_{\cal A} \equiv (x_{\cal A}^M, \theta^{\pm a})$,
where $x^M_{\cal A} \equiv x^M + \frac i2
\theta^{+a}(\gamma^M)_{ab}\theta^{-b}$ and $ \theta^{\pm a}=
u^\pm_k\theta^{ak}$. We use the antisymmetric representation for $6D$  $\gamma$-matrices
 \bea
 (\gamma^M)_{ab} = - (\gamma^M)_{ba}\,, \qquad
 (\widetilde{\gamma}^M)^{ab} = \frac12\varepsilon^{abcd}(\gamma^M)_{cd}\,,
 \eea
where $\varepsilon^{abcd}$ is the totally antisymmetric tensor. By definition, analytic superfields depend
only on the coordinates $(\zeta, u)$, with  $\zeta \equiv (x^M_{\cal A}\,, \theta^{+a})\,$.

We define the spinor and vector $6D$ derivatives in the analytic basis as
 \bea D^+_a=\partial_{-a}\,,
\qquad D^-_a=-\partial_{+a} -2i\theta^{-b}\partial_{ab}\,, \qquad
\partial_{ab} = \frac12 (\gamma^M)_{ab}\partial_M\,.
 \eea
Also we will need the harmonic derivatives
 \bea
D^{\pm\pm}=\partial^{\pm\pm}+i\theta^{\pm a} \theta^{\pm
b}\partial_{ab} + \theta^{\pm a}\partial_{\mp a}\,, \qquad D^0 =
u^{+i} \frac {\partial}{\partial u^{+i}} - u^{-i} \frac {\partial}{
\partial u^{-i}} + \theta^{+a} \partial_{+ a} - \theta^{-a}
\partial_{- a}\,,
 \eea
where $\partial^{\pm }_a \theta^{\pm b} = \delta^b_a$ and the
partial harmonic derivatives are defined as $\partial^{\pm\pm}
= u^{\pm i} \frac {\partial }{ \partial u^{\mp  i}}\,$ (in the cental basis the latter coincide with the full harmonic derivatives).
The spinor and harmonic derivatives satisfy the algebra
 \bea
\{D^+_a,D^-_b\}=2i\partial_{ab}\,, \qquad [D^{++}, D^{--}] = D^0,
\qquad [D^{\pm\pm},D^{\pm}_a]=0\,, \qquad
[D^{\pm\pm},D^{\mp}_a]=D^\pm_a\,.
 \eea
Finally, the full and analytic superspace integration
measures are defined as
 \be
  d^{14}z \equiv d^6x_{\cal A}\,(D^-)^4(D^+)^4,\qquad d\zeta^{(-4)} \equiv d^6x_{\cal
  A}\,(D^-)^4, \qquad (D^{\pm})^4 = -\frac{1}{24}
\varepsilon^{abcd} D^\pm_a D^\pm_b D^\pm_c D^\pm_d V^{--}.
 \ee

Now we briefly recall, basically following ref. \cite{Bossard:2015dva,Buchbinder:2016url,Buchbinder:2017ozh,Buchbinder:2017gbs,Buchbinder:2017xjb,Buchbinder:2018bhs},
some details of the harmonic superspace
formulation of $6D,\,\cN=(1,0)$ SYM interacting with a
hypermultiplet. The classical action of the theory  has the
form
 \bea
 \label{S0}
&& S_0[V^{++}, q^+] =
\frac{1}{f_0^2}\sum\limits^{\infty}_{n=2} \frac{(-i)^{n}}{n}
\tr \int d^{14}z\, du_1\ldots du_n \frac{V^{++}(z,u_1 ) \ldots
V^{++}(z,u_n ) }{(u^+_1 u^+_2)\ldots (u^+_n u^+_1 )}  \nn \\
&& \qquad\qquad\qquad\qquad\qquad\qquad\qquad\qquad\qquad\qquad\qquad\qquad\qquad
- \int d\zeta^{(-4)} du\, \widetilde{q}^{+} \nabla^{++}q^{+},\qquad
 \eea
where $f_0$ is a dimensionful coupling constant ($[f_0]=m^{-1}$). Here $V^{++}$ is the hermitian analytic gauge connection
taking values in the Lie algebra of the gauge group $G$,
 \bea
V^{++}=(V^{++})^A T^A,  \qquad [T^A, T^B] = i f^{ABC} T^C, \qquad
A,B,C=1,..,d_G\,. \label{Vfirst}
 \eea
Here $f^{ABC}$ are the totally antisymmetric structure constants and
$d_G$ is the dimension of the gauge group.

We assume that the hypermultiplet belongs to an irreducible
representation $R$ of the gauge group $G$. Then the covariant harmonic
derivative $\nabla^{++}$ in eq. \eq{S0} acts on the hypermultiplet as
 \bea
 (\nabla^{++})_m{}^n q^{+}_n = D^{++} q^{+}_m + i (V^{++})^A (T^A)_m{}^n q^{+}_n\,,
 \eea
where the generators of the gauge group $T^A$ satisfy the conditions
 \bea
 \tr(T^A T^B) = T(R) \d^{AB},
\qquad (T^A)_m{}^{l} (T^A)_l{}^{n} =  C(R) \delta_m{}^{n}\,.
 \eea
Here $C(R)$ is the second-order Casimir for the representation $R$ and
$T(R)=C(R) d_R/d_G\,$, with  $d_R$ being the dimension of the
irreducible representation $R$. For the adjoint
representation the generators are written as $(T^C_{\rm Adj})_A{}^B
=i f^{ACB}$ and
 \bea
T({\rm Adj}) = C({\rm Adj}) \equiv C_2\,.
 \eea
The generators of the fundamental representation $T^A_{\rm F}=t^A$ are
normalized in the standard way, $\tr( t^At^B ) = \tfrac12
\delta^{AB}$. Hereafter we omit the representation indices on the hypermultiplet.

The action \eq{S0} is invariant under the gauge transformation
 \bea
 (V^{++})' = e^{i\lambda^A T^A} V^{++} e^{-i\lambda^A T^A} -i e^{i\lambda^A T^A}D^{++} e^{-i\lambda^A T^A},\qquad ( q^{+})' =  e^{i\lambda^A T^A }q^{+},\label{gtr}
 \eea
where $\lambda^A(\zeta, u)$ is a real (with respect to the ``tilde'' conjugation) gauge group parameter.

We also introduce the non-analytic harmonic connection $V^{--}= (V^{--})^A T^A$ \cite{Galperin:2001uw} and
construct the second covariant harmonic derivative $\nb^{--}$
 \bea
 \nabla^{--} = D^{--} + i V^{--}.
 \eea
The superfield $V^{--}$ is a solution of the harmonic
zero-curvature condition
 \bea
 D^{++} V^{--} - D^{--}V^{++} + i[V^{++},V^{--}]=0\,,  \label{zeroc}
 \eea
and its explicit expression  in terms of $V^{++}$ is given by
 \be
V^{--}(z,u)=  \sum\limits^{\infty}_{n=1} (-i)^{n+1  } \int
du_1\ldots du_n\, \frac{V^{++}(z,u_1 )\ldots V^{++}(z,u_n )}{(u^+
u^+_1)(u^+_1 u^+_2)\ldots (u^+_n u^+)}\,.
 \ee
Using the zero-curvature condition \eq{zeroc}, one can derive a
useful relation between variations of the gauge harmonic connections
 \be
\d V^{--} = \f12(\nb^{--})^2\d V^{++} - \f12 \nb^{++}(\nb^{--}\delta
V^{--}) \,. \label{var}
 \ee
The superfield $V^{--}$ can  be used to construct the spinor and vector gauge-covariant
derivatives. In the $\lambda$-frame they read
 \bea
 \nb^+_a = D^+_a, \qquad \nb^-_a = D^-_a + i {\cal A}^-_a,
 \qquad \nb_{ab} = \pr_{ab} + i {\cal A}_{ab}\,, \label{deriv}
 \eea
where $\nb_{ab} = \f12(\g^M)_{ab} \nb_M$ and
$\nb_M=\partial_M-iA_M\,$. The superfield connections in eq. \eq{deriv}
are defined as
 \bea
 {\cal A}^-_a= i D^+_a V^{--}, \qquad {\cal A}_{ab} = \f12
 D^+_a D^+_b V^{--}.
 \eea
The covariant derivatives \eq{deriv} satisfy the algebra
 \begin{equation}
\{\nb_a^+,\nb^-_b\}=2i\nb_{ab}\,,\qquad
[\nb_c^\pm,\nb_{ab}]=\tfrac{i}2\ve_{abcd}W^{\pm\, d},\qquad [\nb_M,
\nb_N] = i F_{MN}\,, \label{alg2}
 \end{equation}
where $W^{a\,\pm}$ is the superfield strength of the gauge multiplet\,,
 \bea
W^{+a} \equiv -\frac{i}{6}\varepsilon^{abcd}D^+_b D^+_c D^+_d V^{--},
\qquad W^{-a} = \nb^{--}W^{+a}.\label{W+Def}
 \eea
Also  we define the Grassmann analytic superfield $F^{++}$
\cite{Bossard:2015dva},
 \begin{equation}
 \label{identity00}
F^{++} \equiv (D^+)^4 V^{--}\,, \qquad \nb^{++} F^{++}=0\,.
 \end{equation}

Using the relation between the variations of the gauge connections
$V^{++}$ and $V^{--}$,  we can derive the classical
equations of motion for the model \eq{S0},
 \bea
 \frac{{\delta}S}{{\delta}(V^{++})^A}=0 \,&\Rightarrow& \,
 (F^{++})^A -2 i f_0\, \tilde{q}\, T^A\, q^{+} = 0\,, \lb{eqmF}
 \\
 \frac{{\delta}S}{{\delta}\tilde{q}^{+}} = 0 \,&\Rightarrow& \,\nabla^{++} q^{+} = 0\,. \label{eqm}
  \eea

Finally note that for the hypermultiplet belonging to
the adjoint representation of the gauge group the action \eq{S0}
possesses an additional $\cN=(0,1)$ supersymmetry \cite{Bossard:2015dva}. In
this case the action \eq{S0} describes $6D$,  $\cN=(1,1)$ SYM theory.

\section{The one-loop effective action}
\label{Section_Effective_Action}

The background superfield method for the model \eq{S0}
was developed in refs. \cite{Buchbinder:2016url,Buchbinder:2017ozh,Buchbinder:2017gbs,Buchbinder:2017xjb,Buchbinder:2018bhs}. In many aspects it is
similar to that for $4D$, $\cN=2$ supersymmetric gauge theories
\cite{Buchbinder:1997ya} (see also the review \cite{Buchbinder:2016wng}). Following this method,
we split the superfields $V^{++}, q^{+}$ into the sum of the
\emph{background} superfields $\bm{V}^{++}, Q^{+}$ and the \emph{quantum}
ones $v^{++}, q^{+}\,$,
 \be
 V^{++}\to \bm{V}^{++} + f_0 v^{++}, \qquad q^{+} \to Q^{+} + q^{+}.
 \ee
The effective action is invariant under the \emph{background} gauge
transformations:
\begin{equation}\label{cltr}
\delta \bm{V}^{++}=-\bm{\nb}^{++}\lambda\,, \quad \delta v^{++}=-i[v^{++},\lambda]\,,
\end{equation}
where
\begin{equation}
\bm{\nabla}^{\pm\pm}\equiv D^{\pm\pm} + i \bm{V}^{\pm\pm}
\end{equation}
are the background harmonic covariant derivatives\footnote{Herewith, the bold letters denote the objects constructed out of the background gauge superfield $\bm{V}^{++}$.}.

We use the gauge-fixing function similar to that in  the $4D$ case \cite{Buchbinder:1997ya,Buchbinder:2016wng},
 \be
\label{gf} {\cal F}^{(+4)}_\tau = D^{++}v^{++}_\tau
=e^{-i\bm{b}}(\bm{\nb}^{++} v^{++})e^{i\bm{b}}=e^{-i\bm{b}}{\cal F}^{(+4)}e^{i\bm{b}}~,
 \ee
where $\bm{b}(z,u)$ is the background bridge superfield (see, e.g., \cite{Galperin:2001uw}). In this paper we will use the more general gauge-fixing term
  \be
S_{\mbox{\scriptsize gf}}[v^{++}, V^{++}] = -\frac{1}{2\xi_0}\tr \int d^{14}z du_1
du_2\,\frac{v_\tau^{++}(1)v_\tau^{++}(2)}{(u^+_1u^+_2)^2}
 + \frac{1}{4\xi_0}\tr \int
d^{14}z du\, v_\tau^{++} (D^{--})^2 v_\tau^{++}. \label{SGF}
 \ee
The action \eq{SGF} includes an arbitrary real parameter $\xi_0$ and depends on the background field $\bm{V}^{++}$
through the background gauge bridge $\bm{b}$, $v_\tau^{++} = e^{-i\bm{b}}
v^{++} e^{i\bm{b}}$.

The one-loop quantum correction $\Gamma^{(1)}[\bm{V}^{++},Q^+]\,$ to the classical action \eq{S0} is given by the
following path integral \cite{Buchbinder:2016url,Buchbinder:1997ya}:
 \be
\exp\Big(i\Gamma^{(1)}[\bm{V}^{++}, Q^+]\Big) =\mbox{Det}^{1/2}\sB \int {\cal
D}v^{++}\,{\cal D}q^+\, {\cal D} b\,{\cal D}c\,{\cal
D}\varphi\, \exp\Big(iS_{2}[v^{++}, q^+, b, c, \varphi,
\bm{V}^{++}, Q^+]\Big),
 \label{Gamma0}
 \ee
where  $\bm{\sB}=\tfrac{1}{2}(D^+)^4(\bm{\nb}^{--})^2$ is the covariant d'Alembertian.
When acting on the analytic superfields, it is reduced to
 \bea
 \bm{\sB}=\eta^{MN} \bm{\nabla}_M \bm{\nabla}_N + \bm{W}^{+a} \bm{\nabla}^{-}_a + \bm{F}^{++} \bm{\nabla}^{--} - \frac{1}{2}(\bm{\nabla}^{--} \bm{F}^{++})\,.
 \label{smile}
 \eea

In the expression \eq{Gamma0} $S_2$ denotes that part of the total action which is quadratic in the quantum superfields.
It includes the classical action \eq{S0} in which the
background-quantum splitting is performed, the gauge-fixing action
\eqref{SGF}, and the actions for the ghost superfields,
  \bea
 S_2 &=& S_{\mbox{\scriptsize gh}} + \frac{1}{2\xi_0} \tr\int d\zeta^{(-4)}du\, v^{++}\bm{\sB} v^{++}+
  \frac{1}{2} \Big(1-\frac{1}{\xi_0}\Big)\tr \int d^{14}z du_1
du_2\,\frac{v^{++}(1)v^{++}(2)}{(u^+_1u^+_2)^2}\qquad\nn \\
 && - \int d\zeta^{(-4)}du\, \tilde{q}^{+}
 \bm{\nb}^{++} q^{+} - i f_0\int d\zeta^{(-4)}du\Big(
  \widetilde{Q}^{+} v^{++} q^{+} + \tilde{q}^{+} v^{++} Q^{+}\Big). \label{S2}
 \eea
The ghost actions are written as
 \bea
 S_{\mbox{\scriptsize gh}}=\frac{1}{2}\tr\int d\zeta^{(-4)}du\,\varphi(\bm{\nb}^{++})^{2}\varphi
 + \tr\int d\zeta^{(-4)}du\, b(\bm{\nb}^{++})^{2} c\,.
 \eea
The superfields $b$, $c$ are the Faddeev-Popov ghosts and
$\varphi$ stands for the Nielsen-Kallosh ghost.

The action $S_2$ (\ref{S2}) contains mixed terms in which both the
quantum superfields $v^{++}$ and $q^{+}$ are present. Following ref. \cite{Buchbinder:2016url}, it is convenient to
diagonalize $S_2$ by means of the special change of the quantum
hypermultiplet variables in the path integral\footnote{A similar shift of variables in the path integral of
non-supersymmetric QED was used in ref. \cite{Ostrovsky:1988jn}. The supersymmetric generalization of this procedure
was applied in refs. \cite{Buchbinder:2006td,Kuzenko:2007cg,Buchbinder:2015swa}, while calculating the one-
and two-loop contributions to effective actions of supersymmetric
gauge theories.},
 \bea
 \label{replac}
 q^{+}(1)= h^{+}(1) - if_0 \int d \zeta^{(-4)}_2 du_2\, G^{(1,1)}(1|2)
 v^{++}(2)\, Q^{+}(2)\,.
 \eea
Here, $h^{+}$ is a set of new independent quantum hypermultiplet
superfields and $G^{(1,1)}(1|2)$ is the hypermultiplet Green function,
\bea
  G^{(1,1)}(\zeta_1,u_1|\zeta_2,u_2) = i\langle
{q}^{+}(\zeta_1,u_1) \widetilde{q}^{+\,}(\zeta_2,u_2)\rangle = \frac{(D^+_1)^4(D^+_2)^4}{\bm{\sB}_1}\f{\d^{14}(z_1-z_2)}{(u^+_1u^+_2)^3}\,.
 \label{GREEN}
  \eea
This Green function is analytic with respect to its both super arguments and satisfies the equation
 \bea
 \bm{\nb}_1^{++}G^{(1,1)}(1|2)&=&\d_A^{(3,1)}(1|2)\,,
  \eea
where $\d_A^{(3,1)}(1|2)$ is the covariantly-analytic delta-function
\cite{Galperin:2001uw}.

After performing the shift \eq{replac}, the action $S_2$ \eq{S2} takes the
diagonal form,
 \bea
 S_2 &=& S_{\mbox{\scriptsize gh}} + \frac{1}{2\xi_0} \tr\int d\zeta^{(-4)}du\, v^{++}\bm{\sB} v^{++}
  + \frac{1}{2} \Big(1-\frac{1}{\xi_0} \Big)\tr \int d^{14}z du_1
du_2\,\frac{v^{++}(1)v^{++}(2)}{(u^+_1u^+_2)^2}\qquad\nn \\
 && -f_0^2\int d\zeta^{(-4)}_1 d\zeta^{(-4)}_2 du_1du_2\, v^{++}_1 \widetilde{Q}^+_1 G^{(1,1)}(1|2) Q^+_2 v^{++}_2
 - \int d\zeta^{(-4)}du\, \widetilde{q}^{+}
 \bm{\nb}^{++} q^{+}.  \label{S22}
 \eea
The action \eq{S22} includes a new term which is
quadratic in the quantum vector superfield $v^{++}$ and contains a
non-local contribution involving the Green function $G^{(1,1)}$.

Integrating over the quantum superfields in the path integral
\eq{Gamma0} with the action \eq{S22}, we find the one-loop
contribution $\Gamma^{(1)}$ to the effective action,
 \bea
 \Gamma[\bm{V}^{++},Q;\xi_0]&=&\frac{i}{2}\mbox{Tr}\ln\Big\{ \frac{1}{\xi_0}\bm{\sB}{}^{AB}
 + \Big(1-\frac{1}{\xi_0}\Big)\d^{AB}\frac{(D_1^+)^4}{(u^+_1u^+_2)^2}
 - 4f_0^2 \widetilde{Q}^{+}_1 (T^A G^{(1,1)}T^B)(1|2) Q^{+}_2\Big\}\qquad
  \nn \\
&& -\,\frac{i}{2}\mbox{Tr}\ln\bm{\sB}  -\frac{i}{2}\mbox{Tr}\ln(\bm{\nb}^{++}_{\rm Adj})^2
 +i\mbox{Tr}\ln\bm{\nb}^{++}_{R}\,. \label{1loop}
 \eea
The subscripts ${\rm Adj}$ and $R$ in \eq{1loop} mean that the corresponding
operators act in the adjoint and $R$ representations of the gauge group. The functional trace $\Tr$ in eq. \eq{1loop} is defined as
\begin{equation}
\Tr {\cal O} = \tr \int d \zeta_1^{(-4)}d \zeta_2^{(-4)}
du_1 du_2 \, \d_{\cal A}^{(q,4-q)}(1|2)\,  {\cal
O}^{(q,4-q)}(1|2)\,.
\end{equation}
In this expression ${\cal O}^{(q,4-q)}(\zeta_1,u_1|\zeta_2,u_2)$ is the  kernel
of an operator acting in the space of the covariantly analytic
superfields with the harmonic $U(1)$ charge $q$, and $ \d_{\cal
A}^{(q,4-q)}(1|2)$ is the corresponding analytic delta-function
\cite{Galperin:2001uw},
 \bea
\d_{\cal A}^{(q,4-q)}=
(D^+_2)^4\d^{14}(z_1-z_2)\d^{(q,-q)}(u_1,u_2)\,, \quad \d^{14}(z_1-z_2)=(\theta^+_1-\theta^+_2)^4(\theta^-_1-\theta^-_2)^4\d^6(x_1-x_2)\,.\label{1.6}
 \eea

The expression \eq{1loop} depends on the background superfields $\bm{V}^{++}$
and $Q^+$. Also, it contains the parameter $\xi_0$ of the gauge-fixing term.
Our further purpose will be to study the divergent part of eq. \eq{1loop} as a function of $\xi_0$, without assuming
{\it \`a priori} any restriction on the background superfields.

\section{Gauge dependence of the one-loop divergences}
\label{Section_Divergences}

First, we consider the part of the effective action which is specified by the last two terms in \eq{1loop},
 \bea
\Delta_1\Gamma^{(1)}[\bm{V}^{++}] = -\frac{i}{2}\mbox{Tr}\ln(\bm{\nb}^{++}_{\rm Adj})^2
  +i\mbox{Tr}\ln\bm{\nb}^{++}_R.
 \label{1loopV}
 \eea
It does not depend on the gauge-fixing parameter $\xi_0$ and has been already analyzed earlier in ref. \cite{Buchbinder:2016url}.
Omitting details of the calculation, we present the final expression for
the divergent part of \eq{1loopV}
 \be
 \Delta_1\Gamma^{(1)}_{\infty}  =  \f{C_2 - T(R)}{3(4\pi)^3 \varepsilon}\,
 \mbox{tr}\int d\zeta^{(-4)} du\, (\bm{F}^{++})^2. \label{div7}
 \ee
Here, $\varepsilon\equiv 6-D \to 0$ and $\bm{F}^{++} = (\bm{F}^{++})^{A} t^A$, with $t^A$ being
the generators of the fundamental representation.

The remaining gauge-dependent part of the one-loop counterterms comes out from
the first two  terms in eq. \eq{1loop}
 \bea
 && \Delta_2\Gamma^{(1)}[\bm{V}^{++}, Q^+; \xi_0]\nonumber\\
 && = \,\f{i}2\mbox{Tr}\ln\Big\{\frac{1}{\xi_0}\bm{\sB}{}^{AB}  + \Big(1-\frac{1}{\xi_0}\Big)\delta^{AB}\frac{(D_1^+)^4}{(u^+_1u^+_2)^2}
 - 4f_0^2 \widetilde{Q}^{+}_1 (T^A G^{(1,1)}T^B)(1|2) Q^{+}_2\Big\} - \frac{i}{2}\mbox{Tr}\ln\bm{\sB}\,. \qquad \label{QFQ1}
 \eea
We start with the calculation of the gauge-dependent divergences in the pure gauge superfield sector. So we switch off the background hypermultiplet, $Q^+=0\,$,
and expand the logarithm up to the first order in the inverse $\bm{\sB}$
operator,
 \bea
 \Delta_{2,\,F^2}\Gamma^{(1)}[\bm{V}^{++};\xi_0] &=& \f{i}2\mbox{Tr}\ln\Big\{ \frac{1}{\xi_0}\bm{\sB}
 + \Big(1-\frac{1}{\xi_0}\Big)\frac{(D_1^+)^4}{(u^+_1 u^+_2)^2} \Big\} -\frac{i}{2}\mbox{Tr}\ln\bm{\sB} \nn \\
 &\to & \f{i}2 (\xi_0-1)\int d\zeta^{(-4)}du\,
 (\bm{\sB}_1{}^{-1})^{AA}\frac{(D_1^+)^4}{(u^+_1u^+_2)^2} \d^{(2,2)}_{\cal
 A}(1|2)\bigg|_{2=1}\,.
  \eea
Then we expand the operator $\bm{\sB}{}^{-1}$ up to the second order in the
harmonic derivative $D^{--}$,
 \bea
(\bm{\sB}_1{}^{-1})^{AA} \to \f{1}{(\partial^2)^3} (f^{ACB}(\bm{F}^{++})^{C})  (f^{BDA}(\bm{F}^{++})^{D}) (D^{--})^2+\ldots,
 \eea
and use the explicit expression \eq{1.6} for the analytic delta function
$\d^{(2,2)}_{\cal  A}(1|2)$. On the next step we use the identity
 \bea
 (D^{+}_1)^4 (D^{+}_2)^4 \delta^{14}(z_1-z_2)\bigg|_{\theta_2=\theta_1} = (u_1^+u_2^+)^4
 \delta^6(x_1-x_2)\,,
 \eea
and pass to the momentum representation, taking the coincident-point limit,
 \bea
 \f{1}{(\partial^2)^3} \delta^6(x_1-x_2)\bigg|_{2=1} =
 \f{i}{(4\pi)^3\varepsilon} + \mbox{finite terms}\,.
 \label{int}
 \eea
As a result, we derive the following expression for the divergent part of $\Delta_{2,\,F^2}\Gamma^{(1)}$:
 \bea
 \Delta_{2,\,F^2}\Gamma^{(1)}_\infty[\bm{V}^{++};\xi_0] = \f{2(\xi_0-1)C_2}{(4\pi)^3\varepsilon}\,\tr \int d\zeta^{(-4)}
 du\,(\bm{F}^{++})^2.
\label{FF1}
 \eea

Now we consider the divergent part of the one-loop effective action which depends on the hypermultiplet $Q^+$,
 \begin{eqnarray}
 && \Delta_{2,\,Q^+}\Gamma^{(1)}[\bm{V}^{++}, Q^+;\xi_0] = \frac{i}{2}\int d\zeta^{(-4)}_1 du_1\,
 \Bigg(\frac{1}{\xi_0}\bm{\sB}
 +\Big(1-\frac{1}{\xi_0}\Big)\frac{(D_1^+)^4}{(u_1^+u_2^+)^2}\Bigg)^{-1\, AB}\nonumber\\
 && \qquad\qquad\qquad\qquad\qquad\qquad\qquad
  \Big(-4f_0^2\, \widetilde{Q}^+_1 (T^B G^{(1,1)}T^A) (1|2) Q^+_2
 \Big)\bigg|_{2=1} \qquad \nn \\
 && = -2if_0^2\int d\zeta^{(-4)}_1 du_1\, \Bigg(
\frac{\xi_0}{\sB} -
(\xi_0-1)\frac{1}{\sB{}^2}\frac{(D_1^+)^4}{(u_1^+u_2^+)^2} \Bigg)^{AB}
 \Big(\widetilde{Q}^+_1 (T^B G^{(1,1)}T^A)(1|2) Q^+_2
 \Big)\bigg|_{2=1}. \nn \\ \label{div10}
  \end{eqnarray}
The divergent contribution to the one-loop effective action, which
contains $ \widetilde{Q}^+F^{++}Q^+\,$, can be found similarly to the case of the minimal gauge. It comes out  from the term
 \begin{eqnarray}
 -2i\xi_0f_0^2\int d\zeta^{(-4)}_1 du_1\,
(\bm{\sB}{}^{-1})^{AB} \Big(\widetilde{Q}^+_1 (T^B G^{(1,1)}T^A)(1|2)
Q^+_2 \Big)\bigg|_{2=1}.
 \end{eqnarray}
The divergent part of this expression was calculated in ref. \cite{Buchbinder:2016url}. Here we omit details of the calculation and present the result,
\begin{equation}
\Delta_{2,\,QFQ}\Gamma^{(1)}_{\infty}= -\f{ 2i \xi_0 f_0^2(C_2-C(R))}{(4\pi)^3
\varepsilon} \int d\zeta^{(-4)}du\,\widetilde{Q}^{+}\bm{F}^{++} Q^{+}.
\label{div8}
 \end{equation}

In the case of non-minimal gauges there appears an additional contribution
to the divergent part of the one-loop effective action. It comes from the second term in the \eq{div10},
 \begin{equation}\label{QQ_Part}
 2if_0^2(\xi_0-1)\int d\zeta^{(-4)}_1 du_1\, \Bigg(
\frac{1}{\bm{\sB}^2}\frac{(D_1^+)^4}{(u_1^+u_2^+)^2} \Bigg)^{AB}
 \Big(\widetilde{Q}^+_1 (T^B G^{(1,1)}T^A)(1|2) Q^+_2
 \Big)\bigg|_{2=1}.
  \end{equation}
To cast it in the explicit form, we firstly use the antisymmetry of the hypermultiplet propagator \eq{GREEN},
 \begin{eqnarray}
 \widetilde{Q}^+_1  Q^+_2 G^{(1,1)}(1|2) = \frac12(\widetilde{Q}^+_1 Q^+_2 -\widetilde{Q}^+_2
 Q^+_1)G^{(1,1)}(1|2)\,.
 \end{eqnarray}
Then we rewrite the Green function $G^{(1,1)}$ as \cite{Kuzenko:2001vc,Kuzenko:2001fe,Kuzenko:2003eb,Kuzenko:2004gn,Kuzenko:2006ek}
\begin{eqnarray}
G^{(1,1)}(1|2) &=&\frac{(D_1^+)^4}{\bm{\sB}}\Big\{ (D_1^-)^4 (u^+_1u^+_2)
- \bm{\Omega}^{--}_1(u^-_1 u^+_2) -4
\bm{\sB}\frac{(u^-_1 u^+_2)^2}{(u^+_1 u^+_2)} \Big\}, \label{OmegaG}
\end{eqnarray}
where
 \bea
 \bm{\Omega}^{--} \equiv i\bm{\nb}^{ab}\bm{\nb}^-_a\bm{\nb}^-_b - \bm{W}^{-a}\bm{\nb}^-_a + \frac{1}{4} (\bm{\nb}^-_a
\bm{W}^{- a})~. \label{O}
 \eea
The divergent contribution comes from the first term in the
propagator \eqref{OmegaG}. Plugging it in eq. (\ref{QQ_Part}), we obtain the expression
 \begin{equation}
 if_0^2 (\xi_0-1)\int d\zeta^{(-4)}_1 du_1\,
 \frac{(D_1^+)^4}{\bm{\sB}^2_{\rm Adj}} \frac{1}{(u_1^+u_2^+)^2}
 \Big\{(\widetilde{Q}^+_1 Q^+_2 -\widetilde{Q}^+_2 Q^+_1)
 \frac{(D_1^+)^4(D_1^-)^4}{\bm{\sB}_{R}}(u^+_1u^+_2)\Big\}\delta^{14}(z_1-z_2)\bigg|_{2=1}.
 \end{equation}
We reconstruct the full superspace measure in this expression by taking away the factor $(D^+)^4$ and then use the identities
 \begin{eqnarray}
 && Q^+_2 = (u^+_1u^+_2)Q^-_1 - (u^-_1u_2^+)Q^+_1, \\
 && (\widetilde{Q}^+_1 Q^+_2 -\widetilde{Q}^+_2 Q^+_1) = (u^+_1u^+_2)
 (\widetilde{Q}^+ Q^- -\widetilde{Q}^- Q^+)\,.
 \end{eqnarray}
As the last step, we get rid of the Grassmann delta-function in the limit of coincident points by making use
of the operators $(D^+)^4$ and $(D^-)^4$ and collect the
third power of the $(\partial^2)^{-1}$ operator acting on the
space-time delta-function \eq{int} to extract the divergent
contribution. Finally, we obtain
 \begin{eqnarray}
  \Delta_{2,\, QQ}\Gamma^{(1)}_{\infty} =  -\frac{f_0^2 (\xi_0-1) C(R)}{(4\pi)^3 \varepsilon}
 \int d^{14} z\,du\, (\widetilde{Q}^+ Q^- -\widetilde{Q}^-  Q^+)\,. \label{div9}
  \end{eqnarray}

Summing up the contributions \eq{div7} \eq{FF1}, \eq{div8} and
\eq{div9}, we can present the total  divergent part of the one-loop effective action in the form
 \begin{eqnarray}
 \Gamma^{(1)}_{\infty}[\bm{V}^{++}, Q^+; \xi_0] &=& \Delta_{1}\Gamma^{(1)}_{\infty} + \Delta_{2,\, F^2}\Gamma^{(1)}_{\infty} + \Delta_{2,\, QFQ}\Gamma^{(1)}_{\infty} + \Delta_{2,\, QQ}\Gamma^{(1)}_{\infty}
 \vphantom{\frac{1}{2}} \nonumber\\
 &=& \frac{1}{(4\pi)^3 \varepsilon}\Big(\frac{1}{3}(C_{2} - T(R)) + 2(\xi_0-1)C_2\Big)\, \tr \int d\zeta^{(-4)} du\, (\bm{F}^{++})^2 \nn \\
 &&  -\, \frac{2i\xi_0\,f_0^2(C_2-C(R))}{(4\pi)^3\varepsilon}
 \int d\zeta^{(-4)} du\, {\widetilde Q}^+ \bm{F}^{++} Q^+\nonumber\\
 &&  -\,\frac{f_0^2 (\xi_0-1)C(R)}{(4\pi)^3 \varepsilon}
 \int d^{14} z\,du\, \Big(\widetilde{Q}^+ Q^- -\widetilde{Q}^-  Q^+\Big).
 \label{answer}
 \end{eqnarray}
Note that for the minimal gauge choice, $\xi_0=1$, this expression reproduces the result of refs. \cite{Buchbinder:2016url,Buchbinder:2017ozh}.

According to the general theorem (see, e.g., \cite{DeWitt:1965jb}),
the gauge dependence should vanish on shell. This condition allows
one to check the correctness of eq. (\ref{answer}). Let us suppose
that the background superfields satisfy the classical equations of
motion \eq{eqm}. Then it is easy to show that all terms containing
the gauge-fixing parameter $\xi_0$ are mutually canceled. To this
end, we note that the superfields $Q^+$ and $ Q^-$ are not
independent on shell, $Q^- = \bm{\nb}^{--} Q^+$. Therefore, taking
into account that $\bm{F}^{++} = (D^+)^4 \bm{V}^{--}$, we obtain
\begin{equation}
\int d^{14}z\, du\, \Big(\widetilde Q^+ Q^- - \widetilde Q^- Q^+\Big) = 2\int d^{14}z\, du \widetilde Q^+ \bm{F}^{++} Q^+.
\end{equation}
Consequently,  the divergent part of the one-loop effective action takes on shell the form
\begin{eqnarray}
 \Gamma^{(1)}_{\infty}\Big|_{\mbox{\scriptsize on shell}} &=& \frac{1}{(4\pi)^3 \varepsilon}\Big(\frac{1}{3}(C_{2} - T(R)) + 2(\xi_0-1)C_2\Big)\, \tr \int d\zeta^{(-4)} du\, (\bm{F}^{++})^2 \nn \\
 &&  -\, \frac{2i f_0^2(\xi_0 C_2-C(R))}{(4\pi)^3\varepsilon}
 \int d\zeta^{(-4)} du\, {\widetilde Q}^+ \bm{F}^{++} Q^+.
 \end{eqnarray}
Next, using the equation of motion for the background gauge superfield, $(\bm{F}^{++})^{A} = 2 i f_0^2 \widetilde Q^+ T^A Q^+\,,$ we obtain that on shell the whole
gauge dependence disappears,
\begin{eqnarray}
 \Gamma^{(1)}_{\infty}\Big|_{\mbox{\scriptsize on shell}} &=& \frac{(C_{2} - T(R))}{3(4\pi)^3 \varepsilon}\, \tr \int d\zeta^{(-4)} du\, (\bm{F}^{++})^2  \nn \\
&&  -\, \frac{2i f_0^2(C_2-C(R))}{(4\pi)^3\varepsilon}
 \int d\zeta^{(-4)} du\, {\widetilde Q}^+ \bm{F}^{++} Q^+. \label{On Shell}
\end{eqnarray}

Finally, we note that for the hypermultiplet in the adjoint representation the on-shell result (\ref{On Shell}) evidently vanishes, while the off-shell result (\ref{answer}) does not,
 \begin{equation}
 \Gamma^{(1)}_{\infty}\Big|_{\cN=(1,1)\, \mbox{\scriptsize SYM}} = \frac{2(\xi_0-1)C_2}{(4\pi)^3 \varepsilon}\, \tr \int d\zeta^{(-4)} du\, (\bm{F}^{++})^2
  -\frac{f_0^2 (\xi_0-1)C_2}{(4\pi)^3 \varepsilon}  \int d^{14} z\,du\, \Big(\widetilde{Q}^+ Q^- -\widetilde{Q}^-  Q^+\Big).
 \end{equation}
Thus we come to the conclusion that the one-loop divergences in $\cN=(1,1)$ SYM theory at arbitrary $\xi_0$ vanish only on shell.

\section{Summary}
%\hspace*{\parindent}
\label{Summary}
We considered the general six-dimensional ${\cal N}=(1,0)$
supersymmetric gauge theory in the harmonic superspace formulation
and studied the dependence of the one-loop divergences on the gauge-fixing parameter. The
theory describes ${\cal N}=(1,0)$ vector gauge multiplet
coupled to the hypermultiplet in an arbitrary representation of the
gauge group. The effective action was constructed by means of
the background supefield method, with making use of the one-parameter family of gauges analogous to the $\xi$-gauges of the non-supersymmetric case.
The divergent part of the one-loop effective action was calculated by the superfield proper-time technique, and the gauge dependence of the
divergences was found in the explicit form.

It was shown that the divergent part of the effective action
in the generic gauge contains an additional contribution as compared to the case of
the minimal gauge. However, the whole gauge dependence of
the divergences vanishes for the background superfields satisfying the
classical equations of motion. This confirms the correctness of the calculations performed and
ensures that the divergent part of the $S$-matrix is gauge-independent.

When the hypermultiplet sits in the adjoint representation,  the
results of this paper yield the one-loop divergences of $\cN=(1,1)$
SYM  theory. While using the minimal gauge, such a theory is
off-shell finite at one-loop \cite{Buchbinder:2016url,Buchbinder:2017ozh,Buchbinder:2018bhs}. In this paper
we demonstrated that the one-loop effective action in the
non-minimal gauge is divergent off shell, and this divergence
vanishes only on shell.

One interesting prospect for the future study is
related to a recent activity on constructing the higher-derivative $6D, \, {\cal N}=(1,0)$ supersymmetric gauge theory and
studying the quantum corrections in this theory \cite{Ivanov:2005qf,Smilga:2016dpe,Casarin:2019aqw}. It was noted that such a theory is
renormalizable (modulo anomalies to be manifested in higher loops) and one-loop counterterms were calculated. Note that these divergences were analyzed
in the component formulation, in which supersymmetry is not manifest. So, it
would be extremely interesting to fulfill the superfield quantum
consideration of this higher-derivative theory and to explore the
corresponding effective action in the harmonic superspace formalism. One can expect
that such a manifestly $6D$, ${\cal N}=(1,0)$ supersymmetric analysis will help to reveal more profound aspects of the structure of quantum corrections in this theory.

\section*{Acknowledgements}
The work was supported by the grant of Russian Science Foundation,
project No. 16-12-10306.

%%%%%%%%%%%%%%%%%%%%%%%%%%%%%%%%%%%%%%%%%%%%%%%%%%%%%%%%%%%%%%%%%%%%

\end{document}